\documentclass[aps,superscriptaddress]{revtex4}
\usepackage{amssymb,amsmath,epsfig}
\usepackage[colorlinks=true, pdfstartview=FitV, linkcolor=blue, citecolor=red, urlcolor=magenta, breaklinks=true]{hyperref}
\begin{document}
\title{Analogue Aharonov-Bohm effect in neo-Newtonian theory}
\author{M. A. Anacleto}
\email{ anacleto@df.ufcg.edu.br}
\affiliation{Departamento de F\'{\i}sica, Universidade Federal de Campina Grande
Caixa Postal 10071, 58429-900 Campina Grande, Para\'{\i}ba, Brazil}
\author{I. G. Salako}
\email{ ines.salako@imsp-uac.org; inessalako@gmail.com}
\affiliation{Ecole de Machinisme Agricole et de Construction M\'ecanique -
Universit\'e d'Agriculture de    K\'etou,  BP:43 K\'etou, B\'enin}
\affiliation{Institut de Math\'ematiques et de Sciences Physiques
(IMSP) 01 BP 613 Porto-Novo, B\'enin}
\author{F. A. Brito}
\email{fabrito@df.ufcg.edu.br}
\affiliation{Departamento de F\'{\i}sica, Universidade Federal de Campina Grande
Caixa Postal 10071, 58429-900 Campina Grande, Para\'{\i}ba, Brazil}
\affiliation{Departamento de F\'isica, Universidade Federal da Para\'iba, Caixa Postal 5008, 58051-970 Jo\~ao Pessoa, Para\'iba, Brazil}
\author{E. Passos}
\email{passos@df.ufcg.edu.br}
\affiliation{Departamento de F\'{\i}sica, Universidade Federal de Campina Grande
Caixa Postal 10071, 58429-900 Campina Grande, Para\'{\i}ba, Brazil}
\affiliation{Instituto de F\'{\i}sica, Universidade Federal do Rio de Janeiro, Caixa Postal 21945, Rio de Janeiro, Brazil}

\begin{abstract} 
We address the issues of the scattering of massless planar scalar waves by an acoustic black hole in neo-Newtonian hydrodynamics. We then compute the differential cross section through the use of the partial wave approach in the neo-Newtonian theory which is a modification of the usual Newtonian theory that correctly incorporates  the effects of pressure. We mainly show that the scattering of planar waves leads to a modified analogue Aharonov-Bohm effect due to a nontrivial response of the parameters defining the equation of state.

\end{abstract}
\maketitle
\pretolerance10000

\section{Introduction}
Since the seminal paper by Unruh \cite{Unruh},  the study of analogue models 
of gravity~\cite{MV, Volovik, others} has been an important field to investigate the Hawking radiation as well as to improve the theoretical understanding of quantum gravity. There exists several examples of such analogues, as for example, gravity wave~\cite{RS}, water~\cite{Mathis}, slow light~\cite{UL}, optical fiber~\cite{Philbin} and  electromagnetic waveguide~\cite{RSch}. Specially in fluid systems, the propagation of perturbations of the fluid has been analyzed in many analog models of 
acoustic black holes, such as the models of superfluid helium II~\cite{Novello}, atomic Bose-Einstein condensates~\cite{Garay,OL} and one-dimensional Fermi degenerate noninteracting gas~\cite{SG} that were proposed to create an acoustic black hole geometry in
the laboratory. In addition, the study of a relativistic version of acoustic black holes was presented in~\cite{Xian}.



In the study of analogous models, in general, a classical approach is applied, such as a Newtonian treatment. 
However, the standard Newtonian approach is valid only for pressureless fluids.
In spite of this, in a Newtonian framework, the 
effect of pressure may appropriately be introduced into the dynamics. This is called neo-Newtonian theory which is a modification of the usual Newtonian theory by correctly considering the effects of pressure. {Since pressure is usually an easily adjustable parameter, which may play the role of an external field, the neo-Newtonian theory might provide an interesting way to text analog effects, such as the Aharonov-Bohm (AB) effect due to an acoustic geometry of a vortex in the fluid. As we shall show later, this indeed happens. The AB phase shift which depends on the pressure is analogous to a magnetic flux.}
McCrea in~\cite{McCrea} deduced the neo-Newtonian equations that were later refined in~\cite{Harrison}. 
In addition, in \cite{Lima1997} was obtained a final expression for the equation of fluid considering a perturbative treatment of neo-Newtonian equations (see also \cite{RRRR2003, velten}).
The authors in \cite{Fabris2013} studied acoustic black holes in the framework of neo-Newtonian hydrodynamics  and in \cite{Salako:2015tja} was analyzed the effect of neo-Newtonian hydrodynamics on the superresonance phenomenon.

{ The AB effect~\cite{Bohm} has been used to address several issues in planar physics.  
This is an effect that essentially engenders the scattering of charged particles
by a flux tube. It  has been experimentally confirmed by Tonomura~\cite{RGC} --- for a review see~\cite{Peskin}. 
The effect can also be simulated in  quantum field theory as for example by using a
nonrelativistic field theory for bosonic particles which interact with a Chern-Simons field~\cite{BL}.  }
More specifically,  in~\cite{An} was investigated the AB effect  due to noncommutative spacetime 
in the context of quantum field theory.
{Furthermore,  several other analogues of the AB effect were found in gravitation \cite{FV}, fluid dynamics
\cite{CL}, optics \cite{NNK} and Bose-Einstein condensates \cite{LO}.}
In the context of quantum mechanics the noncommutative AB effect has also been considered in Refs.~\cite{FGLR,Chai}. 
{In particular,  in~\cite{FGLR} has been shown that the noncommutative AB effect, differently from the commutative 
case, develops a non-null cross section for scalar particles scattered by a thin solenoid even with magnetic field assuming discrete values.
Another interesting system was addressed in \cite{Dolan}, where it was shown that  planar waves scattered by a
draining bathtub vortex develops a modified AB effect that has a dependence on two dimensionless parameters related to the circulation and draining rates \cite{Fetter}.  The
effect presents an inherent asymmetry even in the low-frequency regime and leads to novel interference patterns. }
Furthermore, the acoustic black hole metrics obtained from a relativistic fluid
in a noncommutative spacetime~\cite{ABP12} and Lorentz violating 
Abelian Higgs model~\cite{ABP11} have been considered. More recently in~\cite{ABP2012-1}, were extended the analysis made in~\cite{Dolan}  to a Lorentz-violating and noncommutative background \cite{Bazeia:2005tb} which allows to have persistence of phase shifts even if circulation and draining vanish.

{In present study we apply the acoustic black hole metric in neo-Newtonian theory \cite{velten} to obtain the differential cross section for scattered planar waves that leads to a modified AB effect.  As we shall see,
the obtained cross section is similar to that obtained  in \cite{ABP2012-1} for analogue Aharonov-Bohm effect from an idealized draining bathtub vortex and \cite{FGLR} for noncommutative AB effect in quantum mechanics. 
The result shows that pattern fringes persist even in the regime where the parameters related to the circulation vanish. 
In this limit, the neo-Newtonian background forms a conical defect that is also responsible for the analogue AB effect \cite{ABP2012-1}.
 
The paper is organized as follows.
In Sec.~\ref{II} we briefly introduce the acoustic black hole in neo-Newtonian theory.
In Sec.~\ref{AB-gravit} we compute 
the differential cross section due to the scattering  of planar waves that leads to a modified analogue AB effect. Finally in Sec.~\ref{conclu} we present our final considerations.
}

\section{Acoustic black holes in neo-Newtonian hydrodynamics}
\label{II}
In this section we briefly review the neo-Newtonian hydrodynamics and introduce the acoustic black hole metric obtained in \cite{Fabris2013}.

\subsection{neo-Newtonian Hydrodynamics} \label{neonewtonian}

We first consider the standard case of Newtonian equations. 
Thus the basic equations of Newtonian hydrodynamics are
\begin{eqnarray}
\partial_t{\rho}_i + \nabla\cdot(\rho_i\vec{v})  &=& 0, \label{salako100'}
\\
\rho_i \frac{d \vec{v}}{dt} \equiv  \rho_i [\dot{\vec{v}}+
(\vec{v}\cdot\nabla)\vec{v} ] &=& -\nabla p, 
\label{salako200'}
\end{eqnarray}
where $\rho_i$ is the fluid density, $p$ is the pressure and $\vec v$ the {flow/fluid velocity}. 
In this point gravitational interaction is coupled into Euler's equation (\ref{salako200'}) as
\begin{equation}\label{euler1} \dot{\vec{v}}+
(\vec{v}\cdot\nabla)\vec{v}  = - \frac{ \nabla p}{\rho_i} - \nabla \Psi ,
\end{equation}
where
\begin{equation}
\label{poisson1}\nabla^{2} \Psi = 4 \pi G \rho_g,
\end{equation}
and  $ \rho_g $ is the gravitational mass density. 
The expression (\ref{salako100'}) is the continuity equation and (\ref{euler1}) is the Euler equation  modified due to gravitational interaction.

\subsection{Including Pressure}
In neo-Newtonian formalism first we redefine the concept of inertial mass density and gravitational mass density as follows:  
\begin{equation}\label{rhoi}
\rho_i \rightarrow \rho + p, \quad \rho_g \rightarrow \rho + 3p.
\end{equation}
Thus, the neo-Newtonian equations are given by~\cite{ademir,rrrr,velten}
\begin{equation}
\partial_{t} \rho_i + \nabla\cdot(\rho_i\vec v) + p \nabla\cdot\vec{v} = 0\;. 
\label{salako1bis}
\end{equation}
\begin{equation}\label{salako2}
\dot{ \vec{v}} + (\vec{v} \cdot  \nabla )\vec{v} = - \nabla \Psi   - \frac{ \nabla  p}{\rho + p}\,,
\end{equation}
\begin{equation}
\label{salako3}\nabla^{2} \Psi = 4 \pi G \left(\rho + 3 p\right).
\end{equation}
In~\cite{McCrea} this result has been generalized in the presence of pressure.
Moreover, in~\cite{Harrison} this approach has been
modified and leads to neo-Newtonian cosmology.
Note that when $p = 0$ the Newtonian equations are obtained. 

\subsection{Acoustic black holes}

Let us now consider the fluid is barotropic, i.e. $ p=p(\rho)$, inviscid and irrotational where the equation of state $p = k \rho^{n}$, with $k$ and $n$ constants. We write the fluid velocity as $ \vec{v}=-\nabla\psi $ where $ \psi $ is the velocity potential.
Thus, {disregarding the gravitational contribution, we linearise the equations (\ref{salako1bis}) and (\ref{salako2})} by perturbing $ \rho $, $ \vec{v} $ and $ \psi $ as follows:
\begin{eqnarray}
 \rho &=& \rho_{0} + \varepsilon \rho_{1} + 0(\varepsilon^{2})\;, \\
 \rho^{n} &=& \left[\rho_{0} + \varepsilon \rho_{1}+ 0(\varepsilon^{2})\right]^{n} \approx
 \rho^{n}_{0} + n \varepsilon \rho^{n-1}_{0} \rho_{1} + ... \;,
 \\
 \vec{v} &=& \vec{v}_{0} + \varepsilon \vec{v}_{1} + 0(\varepsilon^{2}),
 \\
\psi &=& \psi_{0} + \varepsilon\psi+ 0(\varepsilon^{2})\;, \label{salako80'}
 \end{eqnarray}
{where $\rho$ is the fluid density, $p$ its pressure and $\vec v$ its {flow/fluid velocity}.
Thus, the wave equation becomes
\begin{eqnarray}
&-&\partial_{t} \Big\{c_{s}^{-2}\rho_{0}\Big[\partial_{t}\psi +
\Big(\frac{1}{2} + \frac{\gamma}{2}\Big)\vec v_{0}.\nabla
\psi\Big]\Big\} +\nabla\cdot\Big\{- c_{s}^{-2}\rho_{0}
\vec{v_{0}}\Big[\Big(\frac{1}{2} +
\frac{\gamma}{2}\Big)\partial_{t}\psi 
+ \gamma\vec{v_{0}}.\nabla \psi\Big]+ \rho_{0}\nabla\psi\Big\}=0,
\label{salako12bis}
\end{eqnarray}
that  can be given as
\begin{eqnarray}
\label{eqf}
\partial_{\mu} (f^{\mu\nu} \partial_{\nu} \psi) = 0.
\end{eqnarray}
The Eq. (\ref{eqf})  can also be rewritten as the Klein-Gordon equation for a massless scalar field in a curved (2+1)-dimensional spacetime as follows~\cite{velten}
\begin{eqnarray}
 \frac{1}{\sqrt{-g}} \partial_{\mu} ( \sqrt{-g} g^{\mu\nu} \partial_{\nu} \psi) = 0,   \label{salako14'}
\end{eqnarray}
where
\begin{equation}
f^{\mu\nu}=\sqrt{-g}g^{\mu\nu}= \frac{\rho_{0}}{ c^2_{s}} \left[
\begin{array}{ccc}
-1  & \quad - \frac{(1+\gamma)}{2}v_x & \quad - \frac{(1+\gamma)}{2}v_y \\
\\
- \frac{(1+\gamma)}{2}v_x  &  \quad   c^2_s - \gamma {v}_x^2  & \quad -\gamma {v}_x {v}_y \\
\\
- \frac{(1+\gamma)}{2}v_y  & \quad  -\gamma {v}_x {v}_y  & \quad  c^2_s - \gamma {v}_y^2
\end{array}
\right].
\end{equation}
So in terms of the inverse of $ g^{\mu\nu} $ we obtain the effective (acoustic) metric given in the form
\begin{equation}
g_{\mu\nu}= \sqrt{\frac{\rho_0}{c^2_s+v^2\frac{(\gamma-1)^2}{4}}} 
\left[
\begin{array}{ccc}
-(c^2_s -\gamma v^2)  & \quad - \frac{(1+\gamma)}{2}v_x & \quad - \frac{(1+\gamma)}{2}v_y \\
\\
- \frac{(1+\gamma)}{2}v_x  & \quad   1 +\frac{(\gamma-1)^2}{4c^2_s} v_y^2  &\quad  
-\frac{(\gamma-1)^2}{4c^2_s} v_x v_y \\
\\
- \frac{(1+\gamma)}{2}v_y & \quad -\frac{(\gamma-1)^2}{4c^2_s} v_x v_y   & \quad 1 + \frac{(\gamma-1)^2}{4c^2_s} v_x^2
\end{array}
\right].
\end{equation}
The effective line element can be written as
\begin{eqnarray}
ds^2=\sqrt{\frac{\rho_0}{c^2_s+v^2\frac{(\gamma-1)^2}{4}}}
\left[- (c^2_s -\gamma v^2)dt^2-(1+\gamma)(\vec{v}\cdot d\vec{r})dt +d\vec{r}^2
+\frac{(\gamma-1)^2}{4c^2_s}\left(v_y dx-v_x dy\right)^2\right].
\end{eqnarray}
In polar coordinates ($ \vec{v}=v_r\hat{r}+v_{\phi}\hat{\phi} $ and $ d\vec{r}=dr\hat{r}+rd\phi\hat{\phi} $) we have
\begin{eqnarray}
\label{salako15''} 
ds^2&=&
\tilde{\rho}\left[- \left[c^2_s -\gamma (v^2_r+v_{\phi}^2)\right]dt^2-(1+\gamma)(v_r dr+v_{\phi}rd\phi)dt +(dr^2+r^2d\phi^2)
+\frac{(\gamma-1)^2}{4c^2_s}(v_{\phi} dr-v_{r}rd\phi)^2\right],
\end{eqnarray}
where $ \tilde{\rho}=\sqrt{\rho_0} \left[c^2_s+(v^2_r+v_{\phi}^2)\frac{(\gamma-1)^2}{4}\right]^{-1/2} $ and  $ \gamma=1+kn\rho_0^{n-1} $.
At this point it is appropriate to apply the following coordinate transformations
\begin{eqnarray}
d\tau=dt +\frac{(1+\gamma)v_r dr}{2(c^{2}_s-\gamma v^2_r)}, \quad\quad
d\varphi=d\phi +\frac{\gamma(1+\gamma)v_r v_{\phi}dr}{r(c^{2}_s-\gamma v^2_r)}.
\end{eqnarray}
In this way the line element can be written as
\begin{eqnarray}
\label{eq-kn} 
 ds^{2} &=& \tilde{\rho}
 \Bigg \{- \Big[c^2_s- \gamma(v^2_r+v^2_{\phi})  \Big] d\tau^2  + 
 \frac{c^2_s\left[1+(v^2_{r}+v^2_{\phi}) \left(\frac{\gamma -1}{2c_s}\right)^2 \right]}{(c^2_s-\gamma\,v^2_r)} dr^2
 -v_{\phi} (1+\gamma)\, r d\tau d\varphi 
 \nonumber\\
 &+ &
 r^2\Big[1 + v^2_r \left(\frac{\gamma -1}{2c_s}\right)^2\Big] d\varphi^2 \Bigg\}.
\end{eqnarray}
Now considering a static and position independent density, the {flow/fluid velocity} is given by
\begin{eqnarray}
\vec{v} = \frac{A}{r} \hat r + \frac{B}{r} \hat\phi,
\label{salako17''}
\end{eqnarray}
which is a solution obtained from the continuity equation (\ref{salako100'}) and the velocity potential is
\begin{eqnarray}
\label{salako18'}
 \psi(r,\phi) = - A \ln r - B \phi.    
\end{eqnarray}
Thus, considering $ c_s=1 $ and substituting (\ref{salako17''})
into the metric (\ref{eq-kn}) we obtain, up to an irrelevant
position-independent factor, the acoustic black hole in neo-Newtonian theory  which is given by~\cite{Salako:2015tja}
\begin{eqnarray}
\label{m-ab-nn}
ds^2&=&\beta_1\left[-\left(1-\frac{r^2_e}{r^2}\right)d\tau^2+(1+\beta_2)\left(1-\frac{r_h^2}{r^2}\right)^{-1} dr^2-\frac{2B\beta_3}{r}rd\tau d\varphi+\Big(1+\frac{\beta_4}{r^2} \Big) r^2d\varphi^2\right].
\end{eqnarray}
where 
\begin{eqnarray}\label{betas}
\beta_1 &=&\left(1+\beta_2\right)^{-1/2},
\quad\quad
\beta_2 =\frac{r^2_e}{r^2}\left(\frac{\gamma-1}{2}\right)^2,
\\
\beta_3 &=&\frac{(1+\gamma)}{2}, \quad\quad \beta_4=\left(\frac{A(\gamma-1)}{2}\right)^2,
\end{eqnarray}
being  $ r_e $ the radius of ergo-region and $r_h$ the event horizon, i.e.,
\begin{eqnarray}
r_e=\sqrt{\gamma(A^2+B^2)},  \quad\quad r_h=\sqrt{\gamma}\vert A\vert.
\end{eqnarray}
Thus, the metric (\ref{m-ab-nn}) can be now written in the form
\begin{eqnarray}
g_{\mu\nu}=\beta_1\left[\begin{array}{clcl}
-f &\quad\quad\quad 0& -\frac{B\beta_3}{r}\\
0 & \quad (1+\beta_2){\cal Q}^{-1}& 0\\
-\frac{B\beta_3}{r} &\quad\quad\quad 0 & \left( 1+\frac{\beta_4}{r^2} \right)
\end{array}\right],
\end{eqnarray}
and the inverse of the $g_{\mu\nu}$ is
\begin{eqnarray}
\label{metrinv}
g^{\mu\nu}=\frac{\beta_1(1+\beta_2)}{-g}\left[\begin{array}{clcl}
-\Big(1+\frac{\beta_4}{r^2}\Big){\cal Q}^{-1} &\quad\quad 0& \quad\quad-\frac{B\beta_3}{r{\cal Q}}\\
0 & \quad \frac{-g{\cal Q}}{(1+\beta_2)^2}&\quad\quad 0\\
-\frac{B\beta_3}{r{\cal Q}} &\quad\quad 0 & \quad\quad \frac{f}{{\cal Q}}
\end{array}\right],
\end{eqnarray}
where }
\begin{eqnarray}
f&=&1-\frac{r_e^2}{r^2}, \quad\quad {\cal Q}=1-\frac{r_h^2}{r^2},
\\
-g&=&\frac{(1+\beta_2)}{{\cal Q} }\left[\left(1+\frac{\beta_4}{r^2}\right)f
+\frac{B^2\beta_3^2}{r^2}\right].
\end{eqnarray}
The next step is to consider the Klein-Gordon equation in the background (\ref{m-ab-nn})
\begin{eqnarray}\label{KG-19-08}
\frac{1}{\sqrt{-g}}\partial_{\mu}(\sqrt{-g}g^{\mu\nu}\partial_{\nu})\psi=0,
\end{eqnarray}
with the purpose to study the AB effect. So we can do the following variable separation in the above equation
\begin{eqnarray}
\psi(t,r,\phi)=R(r)e^{i(\omega t-m\phi)}.
\end{eqnarray}
In this way we obtain the following differential equation for the radial function $R(r)$
\begin{eqnarray}
\label{EQKG}
&&\left[\Big(1+\frac{\beta_4}{r^2}\Big)\omega^2-\frac{2B\beta_3 m\omega}{r^2}-\frac{m^2f}{r^2}\right]
\frac{(1+\beta_2)R(r)}{(-g){\cal Q}}
+\frac{1}{r\sqrt{-g}}\frac{d}{dr}\left[r\sqrt{-g}(1+\beta_2)^{-1}{\cal Q}\frac{d}{dr}\right]R(r)=0.
\end{eqnarray}
We can rewrite the equation (\ref{EQKG}) as
\begin{eqnarray}
\label{EQKGbtz}
&&\left[\Big(1+\frac{\beta_4}{r^2}\Big)\omega^2-\frac{2B\beta_3 m\omega}{r^2}-\frac{m^2f}{r^2}\right]R(r)
+\frac{{\cal F}(r)}{r}\frac{d}{dr}\left[r{\cal F}(r)\frac{d}{dr}\right]R(r)=0,
\end{eqnarray}
where $ {\cal F}(r)=\sqrt{-g}(1+\beta_2)^{-1}{\cal Q}(r) $.  
At this point we introduce the coordinated $\varrho$ using the following equation~\cite{Gamboa}
\begin{eqnarray}
\frac{d}{d{\varrho}}={\cal F}(r)\frac{d}{dr},
\end{eqnarray}
and now introducing the new radial function $G(\varrho)=r^{1/2}R(r)$, we can obtain the following modified radial equation
\begin{eqnarray}
\label{EG}
\frac{d^2G(\varrho)}{d\varrho}+\left\{\left[\left(1+\frac{\beta_4}{r^2}\right)^{1/2}\omega
-\frac{B\beta_3 m}{r^2}\left(1+\frac{\beta_4}{r^2}\right)^{-1/2} \right]^2-V(r)\right\}G(\varrho)=0, 
\end{eqnarray}
and the potential $V(r)$ is given by~\cite{Dolan, ABP2012-1}
\begin{eqnarray}
\label{potv}
V(r)=\frac{{\cal F}(r)}{4r^2}\left[\frac{4m^2f(r)}{{\cal F}(r)}-{\cal F}(r)
+\frac{4B^2\beta_3^2m^2}{r^2{\cal F}(r)}\left(1+\frac{\beta_4}{r^2}\right)^{-1}+2r\frac{d{\cal F}(r)}{dr}\right].
\end{eqnarray}
{ We note that the equation (\ref{potv}) does not satisfy the asymptotic behavior $V(r)\rightarrow 0 $ as 
$ r\rightarrow\infty $.  }
Again we introduce a new function $X(r)={\cal F}(r)^{1/2}G(\varrho)$ and we rewrite equation (\ref{EG}) that reads
\begin{eqnarray}
\label{eqX}
\frac{d^2X(r)}{dr^{2}}&+& \left[-\frac{1}{2{\cal F}(r)}\frac{d^2{\cal F}(r)}{dr^2}+\frac{1}{4{\cal F}^2(r)}\left(\frac{d{\cal F}(r)}{dr} \right)^2\right]X(r)
\nonumber\\
&+&\left\{ \left[\left(1+\frac{\beta_4}{r^2}\right)^{1/2}\omega
-\frac{B\beta_3m}{r^2}\left(1+\frac{\beta_4}{r^2}\right)^{-1/2} \right]^2-V(r)\right\}
\frac{X(r)}{{\cal F}^2(r)}=0, 
\end{eqnarray}
where 
\begin{equation}
\frac{d{\cal F}(r)}{dr}=\frac{2{\cal F}}{r}\left[-\frac{1}{1+\beta_2}+\frac{1}{{\cal Q}} \right]
-\frac{{\cal F}}{r^3}\left[\frac{r_h^2}{{\cal Q}}+\frac{(\gamma-1)^2r_e^2}{4(1+\beta_2)}
-\frac{\beta_3^2 B^2+r_e^2\left(1+\frac{\beta_4}{r^2}\right)-\beta_4 f}
{I_3}\right],
\end{equation}
and 
\begin{eqnarray}
\frac{d{^2\cal F}(r)}{dr^2}&=&{\cal F}\left\{\frac{r_e^2(\gamma-1)^2}{2(1+\beta_2) r^3}\left[\frac{r_e^2(\gamma-1)^2}{(1+\beta_2) r^3}+  I_1+ \frac{6r_h^2}{{\cal Q}r^3} 
-\frac{3}{2r} \right]-\frac{1}{4}(I_1+I_2)^2+\frac{2r_h^2(I_1+I_2)}{(1+\beta_2)r^3}
+\frac{4r_h^4}{{\cal Q}r^6}\right.
\nonumber\\
&-&\left. \frac{3r_h^2}{{\cal Q}r^4} -\frac{4r_h^2 I_2}{2{\cal Q}r^2} +
\frac{1}{I_{3}r^4}\left[3\beta_{3}^{2}B^2 +3r_e^2\left( 1+\frac{\beta_4}{r^2}\right) 
-\frac{4\beta_4r_e^2}{r^2}+3\beta_4{\cal Q}\right]\right\},
\end{eqnarray}
being, 
\begin{eqnarray}
I_1=-\frac{2r_h^2}{{\cal Q}r^3}-\frac{r_e^2(\gamma-1)^2}{2(1+\beta_2)r^3},\quad
I_2=\frac{2}{I_3r^3}\left[r_e^2\left(1+\frac{\beta_4}{r^2}\right)-\beta_3^2B^2-\beta_4{\cal Q}  \right],
\quad
I_3=\left(1+\frac{\beta_4}{r^2}\right)f+\frac{\beta_3^2B^2}{r^2}.
\end{eqnarray}
At this point it is convenient to write  (\ref{eqX}) as a power series in $1/r$, 
\begin{eqnarray}
\label{eqv}
\frac{d^2X(r)}{dr^{2}}
+\left[\omega^2-\frac{4\alpha\widetilde{m}^2-1}{4r^2}+U(r)\right]X(r)=0, 
\end{eqnarray}
where
\begin{eqnarray}
U(r)=\frac{1}{4\omega^2 r^4}O(a^2,b^2)+\frac{1}{64\omega^4r^6}O(a^4,b^4),
\end{eqnarray}
and we have defined 
\begin{eqnarray}\label{alphas}
\widetilde{m}^2&=&\frac{m^2}{\alpha}+2am-2b^2+\frac{\beta_5^2a^2}{\alpha\beta_3^2}, \quad
a=\frac{\beta_3}{\alpha}{\omega} B, \quad b=\frac{\beta_3}{\alpha}{\omega} A, \quad 
\nonumber\\
\beta_5&=&\left[\beta_3^2-\gamma\left(1+\frac{(\gamma-1)}{4}\right)\right]^{1/2}, 
\quad
\alpha=\gamma\left[1+\frac{(\gamma-1)^2}{8} \right].
\end{eqnarray}
Here $a$ is the circulation parameter and $b$ is the parameter that describes the draining.
{The total potential $V(r)\equiv-(4\alpha\widetilde{m}^2-1)/r^2+U(r)$ in equation (\ref{eqv}) satisfies the expected asymptotic behavior $V(r)\rightarrow 0$ as $ r\rightarrow\infty $. 
Thus, we can now apply the usual techniques to find Aharonov-Bohm effect.}
 
 \section{Analogue Aharonov-Bohm effect}
\label{AB-gravit}
{In this section we will study the analogue Aharonov-Bohm effect in neo-Newtonian theory. For this purpose we consider the scattering of a monochromatic planar wave of frequency $\omega$ as~\cite{jack}}
\begin{eqnarray}
\psi(t,r,\phi)=e^{-i\omega t}\sum_{m=-\infty}^{\infty}R_{m}(r) e^{im\phi}/\sqrt{r},
\end{eqnarray}
with the function $\psi$ written in the form
\begin{eqnarray}
\psi(t,r,\phi)\sim e^{-i\omega t}(e^{i\omega x}+f_{\omega}(\phi)e^{i\omega r}/\sqrt{r}),
\end{eqnarray}
where $e^{i\omega x}=\sum_{m=-\infty}^{\infty}i^mJ_{m}(\omega r) e^{im\phi}$ and $J_{m}(\omega r)$ 
is a Bessel function of the first kind. 
In this case using the representation of partial waves scattering amplitude $f_{\omega}(\phi)$ reads
\begin{eqnarray}
f_{\omega}(\phi)= \sqrt{\frac{1}{2i\pi\omega}}\sum_{m=-\infty}^{\infty}(e^{2i\delta_{m}}-1) e^{im\phi}.
\end{eqnarray}
Now we can calculate $\delta_{m} $ by applying the following expression
\begin{eqnarray}
\delta_{m}\approx \frac{\pi}{2} (m-\tilde{m})+\frac{\pi}{2}\int^{\infty}_{0}r[J_{\tilde{m}}({\omega}r)]^2U(r)dr,
\end{eqnarray}
 and also using $|m|\gg\sqrt{a^2+b^2}$, we obtain~\cite{Dolan,ABP2012-1}
\begin{eqnarray}
\label{fase}
\delta_{m}\cong \frac{\pi}{2}\left(1-\alpha^{-1/2}\right)m\frac{m}{|m|}
-\frac{\pi\alpha^{1/2} a}{2}\frac{m}{|m|}+O(m^{-1},a^2,b^2).
\end{eqnarray}
{ Therefore, in this case we have obtained the differential scattering cross section  
restricted to small angles $\phi$ as follows 
}
\begin{eqnarray}
\label{sc}
\frac{d\sigma}{d\phi}&=&|f_{\omega}(\phi)|^2
\cong \frac{\alpha\pi^2 a^2}{2\pi\omega}
\left[\left(\frac{4}{\phi^2}-\frac{2}{3}+\frac{\phi^2}{60}+O(\phi^3)\right) 
+\left(1-\alpha^{-1/2}\right)\left(- \frac{4\pi}{\phi^3}+\frac{\pi\phi}{60}+O(\phi^3) \right)\right.
\nonumber\\
&+&\left.\left(1-\alpha^{-1/2}\right)^2\left(\frac{4\pi^2}{\phi^4}-\frac{\pi^2}{3\phi^2}-\frac{\pi^2}{45}-\frac{\pi^2 \phi^2}{15120}+O(\phi^3)\right)\right]
\nonumber\\
&+&\frac{\left(1-\alpha^{-1/2}\right)^2}{2\pi\omega}\left(\frac{4\pi^2}{\phi^4}+\frac{2\pi^2}{3\phi^2}+\frac{11\pi^2}{180}+\frac{31\pi^2 \phi^2}{7560}+O(\phi^3)\right).
\end{eqnarray}
Note that when $\alpha=1 $ in (\ref{fase}),  the phase shift is 
\begin{eqnarray}
\delta_{m}\cong -\frac{a\pi}{2}\frac{m}{|m|},
\end{eqnarray}
and for the differential scattering cross section (\ref{sc}) at small angles $\phi$ we have
\begin{eqnarray}
\frac{d\sigma}{d\phi}=|f_{\omega}(\phi)|^2\cong\frac{\pi^2 a^2}{2\pi\omega}\left[\frac{4}{\phi^2}-\frac{2}{3}+\frac{\phi^2}{60}+O(\phi^3)\right],
\end{eqnarray}
so when $\phi \rightarrow 0 $, the result for the differential cross section is  
\begin{eqnarray}
\frac{d\sigma}{d\phi}=\frac{\pi^2 {a}^2}{2\pi\omega}\frac{4}{\phi^2},
\end{eqnarray}
which is the differential scattering cross section for the analogue AB effect due to the metric of an acoustic black hole.

On the other hand when the parameter associated with circulation is null, $ a=0 $, the equation (\ref{sc}) becomes
\begin{eqnarray}
\label{eqa0}
\frac{d\sigma}{d\phi}=|f_{\omega}(\phi)|^2\cong\frac{(1-\alpha^{-1/2})^2}{2\pi\omega}
\left(\frac{4\pi^2}{\phi^4}+\frac{2\pi^2}{3\phi^2}+\frac{11\pi^2}{180}+\frac{31\pi^2 \phi^2}{7560}+O(\phi^3)\right).
\end{eqnarray}
Thus taking $\phi \rightarrow 0 $, the equation (\ref{eqa0}) now becomes
\begin{eqnarray}\label{sigma-a-b-0}
\frac{d\sigma}{d\phi}=\frac{(1-\alpha^{-1/2})^2 }{2\pi\omega}\frac{4\pi^2}{\phi^4}.
\end{eqnarray}
{This shows that we have successfully obtained the analogue AB scattering in neo-Newtonian theory. We mainly found that differently from the usual Aharonov-Bohm effect, the differential
scattering cross section does not vanish even if $a=0$. 
In addition, for $ \alpha=1 $ and $ a\neq 0 $ in (\ref{sc})  the scattering cross section is symmetric under 
$ \phi \rightarrow-\phi $, 
while for $\alpha\neq 1 $ and $ a\neq 0 $ the low-frequency scattering cross section turns out to be asymmetric, with 
}
\begin{eqnarray}
\sigma_\bot=\int_{-\pi}^{\pi}\frac{d\sigma}{d\phi}\sin\phi d\phi
=-\frac{\alpha(1-\alpha^{-1/2})\pi^2 a^2}{\pi\omega},
\end{eqnarray}
which is due to contribution $ 1/\phi^3 $ in equation (\ref{sc}) for the differential cross section. Recall that $\alpha$ has a non-trivial dependence on the parameters of the equation of state of the fluid.

{We mainly notice from Eq.~(\ref{sigma-a-b-0}) that the pressure develops an important activity in the AB effect even in the limit of the circulation $a=\frac{\beta_3}{\alpha}{\omega} B$ and draining $b=\frac{\beta_3}{\alpha}{\omega} A$ parameters vanish or become very small. For the latter case, we can interpret this result for an infinitly thin vortex as follows. }
{ Now, by making the following changes of variables $t=(\beta_3)^{1/2}\tau$ and $\phi=(\beta_3)^{1/2}\varphi$ 
with $\beta_3=(1+\gamma)/2$, the metric (\ref{m-ab-nn}) becomes
\begin{eqnarray}
\label{m-ab-knn}
ds^2&=&\beta_1\left[-\beta^{-1}_3\left(1-\frac{r^2_e}{r^2}\right)dt^2+(1+\beta_2)\left(1-\frac{r_h^2}{r^2}\right)^{-1} dr^2-\frac{2B\beta_3^{-1}}{r}rdt d\phi+\Big(1+\frac{\beta_4}{r^2} \Big) r^2(\beta_3)^{-1}d\phi^2\right].
\end{eqnarray}
In such a limit, assuming $\omega\neq0$ and $\beta_3\neq0$, we have $\beta_4=\beta_2=A = B = r_h = r_e = 0$, $\beta_1=1$ and the metric (\ref{m-ab-knn})  can be written as }
\begin{eqnarray}
ds^2=-(\beta_3)^{-1}dt^2 + dr^2 + r^2(\beta_3)^{-1}d\phi^2,
\end{eqnarray}
{The metric clearly describes a conical space with angle deficit $\delta=2\pi(1-(\beta_3)^{-1/2})$. If we embed this metric in 3+1 dimensions it can be seen as the background of an infinitely thin low spinning `cosmic' string-like defect along the third spatial dimension. Since the equation of state of the present fluid is $p=k\rho^n$ then $\gamma=1+kn\rho_0^{n-1}=1+np_0/\rho_0$ or simply $\gamma=1+c_s^2$, where $p_0$ and $\rho_0$ are the pressure and density of the fluid background, we can find $\delta\sim\frac{\pi}{2}n{p_0}/{\rho_0}$, for $np_0/\rho_0\ll1$. Similarly, from (\ref{alphas}), $\alpha\sim 1+np_0/\rho_0$ and now we can rewrite Eq.~(\ref{sigma-a-b-0}) as follows
\begin{eqnarray}
\frac{d\sigma}{d\phi}\simeq \frac{2\delta^2}{\pi\omega}\frac{1}{\phi^4}.
\end{eqnarray}
This result easily shows how the fluid pressure affects the AB effect that comes from a conical defect in the acoustic geometry due to an infinitely thin vortex. The AB phase shift $\sim \delta\sim c_s^2$ which depends on the pressure is analogous to a magnetic flux. A related reasoning has been considered in graphene physics \cite{vozmediano}. Our study suggests how to measure such an effect through the scattering of quasiparticles (phonons) in the fluid due to infinitely thin filaments in a bulk fluid, playing the role of the infinitely thin slow spinning string-like defects. The pattern fringes could be detected by adjusting the pressure of the fluid background.
}

\section{Conclusions}
\label{conclu}

{ In summary, in the present study we investigated the analogue Aharonov-Bohm effect from the metric of an acoustic black hole in the neo-Newtonian hydrodynamics. To address the issues concerning this effect we considered the scattering of a monochromatic planar wave.
The differential cross section is shown to be qualitatively in agreement with that obtained in~\cite{FGLR} for the AB effect in the context of noncommutative quantum mechanics and in~\cite{ABP2012-1} for an analogue Aharonov-Bohm
effect from an idealized draining bathtub vortex and {\it gravitational} Aharonov-Bohm effect from a noncommutative BTZ black hole. The correction to differential cross section approaches zero in the limit $\alpha\rightarrow 1$ so that no singularities are found.
The result also shows that pattern fringes can appear even if $a$ vanishes, differently from the standard AB effect. It is interesting to notice that the cross section has a non-trivial dependence on the parameters of the equation of state of the fluid, encoded in the parameter $\alpha$. Comparing this result with that obtained in \cite{ABP2012-1} (third reference) for noncommutative BTZ black hole we see that there exists an equivalence between $\alpha$ and the product of the cosmological constant times the noncommutative parameter. Some further investigations on this relationship should be addressed elsewhere. 
}

\acknowledgements
We would like to thank CNPq and CAPES for partial financial support. 
{ Ines G. Salako thanks  IMSP  for hospitality during the elaboration of this
work.}

\end{document}